# How are excellent (highly cited) papers defined in bibliometrics?

# A quantitative analysis of the literature

Lutz Bornmann

Division for Science and Innovation Studies

Administrative Headquarters of the Max Planck Society

Hofgartenstr. 8,

80539 Munich, Germany.

E-mail: bornmann@gv.mpg.de

**Abstract**

As the subject of research excellence has received increasing attention (in science policy) over the last few decades, increasing numbers of bibliometric studies have been published dealing with excellent papers. However, many different methods have been used in these studies to identify excellent papers. The present quantitative analysis of the literature has been carried out in order to acquire an overview of these methods and an indication of an "average" or "most frequent" bibliometric practice. The search in the Web of Science yielded 321 papers dealing with "highly cited", "most cited", "top cited" and "most frequently cited". Of the 321 papers, 16 could not be used in this study. In around 80% of the papers analyzed in this study, a quantitative definition has been provided with which to identify excellent papers. With definitions which relate to an absolute number, either a certain number of top cited papers (58%) or papers with a minimum number of citations are selected (17%). Around 23% worked with percentile rank classes. Over these papers, there is an arithmetic average of the top 7.6% (arithmetic average) or of the top 3% (median). The top 1% is used most frequently in the papers, followed by the top 10%. With the thresholds presented in this study, in future, it will be possible to identify excellent papers based on an "average" or "most frequent" practice among bibliometricians.

# 1    Introduction

Citations are count data, with an, as a rule, skewed distribution over the papers in a publication set: on the one hand there a few highly cited papers and on the other a large quantity of papers which are rarely, if ever, cited. Since the end of the 1990s, bibliometrics has been looking at this small group of highly cited papers particularly closely. There is a "shift from bibliometric impact scores based on average values such as the average impact of all papers published by some unit to be evaluated towards indicators reflecting the top of the citation distribution, such as the number of 'highly cited' or 'top' articles" (van Leeuwen, Visser, Moed, Nederhof, & van Raan, 2003, p. 257). One reason for this shift is undoubtedly that science policy is increasingly interested in scientific excellence given its new public management tools (Aksnes, 2003; Lamont, 2012). "Many countries are moving towards research policies that emphasise excellence; consequently; they develop evaluation systems to identify universities, research groups, and researchers that can be said to be 'excellent'" (Danell, 2011, p. 50).

However, the general public is also showing more interest in ranked lists of journals, highly cited papers or educational institutions (Bar-Ilan, 2008), to find information about research excellence. The SCImago Institutions Ranking (Bornmann, de Moya Anegón, & Leydesdorff, 2012) and the Leiden Ranking (Waltman et al., 2012) have both included an indicator in the analyses which gives information about the proportion of excellent papers from a university. The high level of public and science policy interest in scientific excellence is supported by studies such as Bornmann, de Moya-Anegón, and Leydesdorff (2010): their comprehensive study has shown that highly cited papers in all scientific disciplines are more strongly based on previously highly cited papers than on medium cited papers. In other words, they are able to demonstrate that papers contributing to the scientific progress in a discipline refer to a larger extent on previously important papers than papers contributing little.

Over recent years, a number of studies have been concerned with excellent and/or highly cited papers. For example, Small (2004) surveyed authors of highly cited papers in 22 fields, in order to discover their opinions on why their papers are highly cited. The responses of the authors to the assumed reasons related to the strong interest, the novelty, the utility and the high significance of the research reported on in the papers. According to the findings of Wang (2013) the accuracy of using short time windows for citation impact measurements is low for highly cited papers, "because they have [a] longer citation life and therefore require longer time periods to reveal their full impacts" (p. 858, see also Bornmann, Leydesdorff, & Wang, 2013). Aksnes (2003) identified highly cited papers in a Norwegian study. The results of the study show that they are typically authored "by a large number of scientists, often involving international collaboration. The majority of the papers represent regular journal articles (81%), although review articles (12%) are over-represented compared to the national average … Highly cited papers typically obtain citations from a large number of different journals and from papers representing both close and remote fields" (p. 159). In the study Aksnes (2003) also notes the difficulty that "there are various definitions of what counts as a highly cited article. Basically two different approaches can be identified, involving absolute or relative thresholds" (p. 160). For example, with an absolute threshold, a certain number of citations is used to identify highly cited papers; with a relative threshold a relative measure such as the top 10% most highly cited papers in a discipline or in a publication year is used.

Although one witnesses an increased focus on research excellence within and outside of science it is not yet clear how the range of excellent papers (i.e. "highly cited", "top cited" or "most frequently cited" papers) should be defined (Glänzel & Schubert, 1992; Kostoff, Barth, & Lau, 2008). Up to now, very different methods have been used to do this. That is why one objective of this quantitative analysis of the bibliometric literature is to determine and show the various methods. Furthermore, it is also intended to establish for the frequently used methods which quantitative thresholds are used to identify the excellent papers. For example, how often percentile rank classes (the proportion of top x% papers in a publication set) have been used to identify excellent papers in the past and which thresholds (x) are used most frequently. As meaningful definitions and thresholds are only used by experts in bibliometrics, a quantitative analysis of (empirical) papers in the library and information sciences is carried out in this study. As using a specific bibliometric method to define research excellence is a normative attribution, the definition and thresholds used in real empirical studies (by bibliometricians) are determined.

## 2 Methods

### 2.1 Data

The literature analysed for this study was researched on 25 February 2013 in the Web of Science (Thomson Reuters). In order to obtain those papers (articles, reviews and letters) which have bibliometrically investigated excellent papers, a topic search for "highly cited", "most cited", "top cited" and "most frequently cited" was undertaken in the SCI Expanded and in the SSCI. The designation "excellent" is not usual on the paper level and rarely found in the literature (as "excellent paper"). However, "excellent paper" is used here as a general term for the various designations, such as highly cited and most cited paper. In other words, excellence is used simply as a synonym of high citedness.

The search in the Web of Science was related to all publication years but was limited to the subject category "Information Science & Library Science". This was intended to ensure that the papers were researched in the subject category (journal set) in which bibliometricians usually publish. From bibliometricians, as experts in the analysis of citation data, statements have been hoped on what an excellent paper is in bibliometric terms. Although one cannot assume of every researched paper that it has been written by an expert in bibliometrics. By restricting the search to journals in the area of information science and library science, one can expect that each paper (and therefore also the definition of excellent papers) has been reviewed by a bibliometrician and classified as worthy of publication. The search in the Web of Science yielded 321 papers. Of the 321 papers, 16 could not be used in this study because they were either not or no longer accessible or they were written in a language other than English. As more than one definition of excellent papers was given in some papers (n=44), it was possible to include 365 entries in this study.[1]

## 2.2    Coding the data for quantitative analysis[2]

The data from the papers researched was coded by the author of this study on two separate occasions, in order to correct any errors in the first coding. If there were differences in the coding between the first and second time, the data from the paper was reviewed where necessary.

The following information was recorded for each of the papers:

1) How is an excellent paper defined in bibliometric terms? For example, does one use a database producer' definition or does one select a certain proportion of top cited papers?

2) What discipline does the paper refer to? The information in the papers was assigned to four subject areas (Bornmann, et al., 2010): (i) <u>Life Sciences</u>: Agricultural & Biological Sciences; Biochemistry, Genetics & Molecular Biology; Immunology & Microbiology; Neuroscience; Pharmacology, Toxicology & Pharmaceutics. (ii) <u>Health Sciences</u>: Medicine; Nursing; Veterinary; Dentistry; Health Professions. (iii) <u>Physical Sciences</u>: Chemical Engineering; Chemistry; Computer Science; Earth & Planetary Science; Energy; Engineering; Environmental Science; Materials Science; Mathematics; Physics & Astronomy. (iv) <u>Social Sciences</u>: Arts & Humanities; Business, Management & Accounting;

---

[1] For example, Zhu, Wu, Zheng, and Ma (2004) use both 10 and 20 citations as thresholds to identify excellent papers.
[2] The bibliographic data of the papers (included in the study) and codes assigned to the papers can be downloaded from http://www.lutz-bornmann.de/excellence/support.xlsx

Decision Sciences; Economics, Econometrics and Finance; Psychology; Social Sciences. If a paper was related to more than one subject area, it was categorised under "Multidisciplinary sciences". The 10 papers in which it was not possible to find a reference to a specific subject area were also assigned to the category "Multidisciplinary sciences".

3) For which research unit are results on excellent papers presented in the researched papers (e.g. single scientists or single papers)?

4) Which term is used for an excellent paper (e.g. highly cited, most highly cited, or most frequently cited paper)?

**2.3   Statistical procedures**

The association between two categorical variables is tested using Pearson's chi-square test (Agresti, 2002). This significance test works fine so long as all expected frequencies in a corresponding table are above about 1. Since the result of the test is dependent on sample size and "statistical significance does not mean real life importance" (Conroy, 2002, p. 290), it is the strength of the association that is more interesting and important for interpreting the empirical finding. For calculating strength, one has to employ an additional measure of association; that is, Cramer's V coefficient (Cramér, 1980). According to Kline (2004), Cramer's V "is probably the best known measure of association for contingency tables" (p. 151).

# 3    Results

Table 1 shows how excellent papers were defined bibliometrically for the papers included in this study. The results are shown broken down by subject area. As this breakdown shows, most papers relate to the social sciences (n=112; and here primarily to "library and information science") and to the multidisciplinary sciences (n=110); the life sciences have the fewest papers. The definitions given in the papers were coded as follows: (1) <u>Database</u>: In around 6% of the papers the authors referred to a definition used in a literature database such as the Essential Science Indicators (Thomson Reuters). In the Essential Science Indicators excellent papers are those papers belonging to the 1% most cited papers by subject area and year of publication in the last 10 years. Other sources of data are Thomson Reuters' ISI Highly Cited (http://www.highlycited.com) and Thomson Reuters' Citation Classics (http://archive.sciencewatch.com/dr/cc/). (2) <u>Qualitative</u>: In around 20% of the papers, excellent papers are selected, but no specific definition <u>given</u>. In some of these studies, the authors select excellent papers without <u>using</u> a specified definition.

(3) <u>Quantitative</u>: A precise quantitative definition to identify excellent papers is given in around 30% of the papers (see Table 1). The definitions which authors refer to specifically are given later in this section (for example, for one of the definitions "excellent" refers to a certain proportion of papers which belong to the 10% most cited papers in their subject area and publication year) (4) <u>Top number</u>: In almost half of the papers (45.24%) the author refers to a certain number of top cited papers, such as the most highly cited paper or the ten most cited papers in a publication set. The results of analyses relevant to the specific number of top cited papers used in the researched papers to identify excellent papers will be presented later in this section.

The comparison of subject areas in Table 1 used to identify the ways to identify excellent papers for a study in different subject areas. The table shows that the use of a database definition is particularly popular in the life sciences (17.39%) but hardly happens at all in the health (0%) and social sciences (0.89%). Compared to the other subject areas, the physical sciences rarely do without a quantitative definition of excellent papers (qualitative = 11.11%) and unlike the other subject areas, use them most frequently (quantitative = 36.11%). If a researched paper refers to more than one subject area (multidisciplinary sciences), it uses a certain number of top cited papers to designate excellent papers more rarely (30%) than if the paper relates to a specific subject area (from 47.22% for physical sciences and 53.57% for social sciences). As citation impact is dependent on the subject area (Bornmann & Marx, 2013) and therefore absolute thresholds should only be used in a study of a single subject area, the significantly more frequent use of these definitions in a study in one subject area than in a study in several subject areas indicates an informed use in the researched papers.

Table 2 shows which unit under study was referenced in the papers concerned with excellent papers. As the figures in the table show, the majority of studies referred to individual excellent papers (64.84%). Around a third of the papers included in this study were about scientists (15.56%) or journals (14.41%).

Table 3 shows the different designations used in the papers to denote excellence and the dependence of the designations on the definitions of excellent papers (see Table 1). As more than one designation is used in some papers which were quantitatively analysed in this study a distinction is made between the number of papers and the number of entries in the papers in the lower section of the table. As the results in the table (see Total column) indicate around a half of the papers (51.59%) use the designation "highly cited" and in another third (32.56%) "most cited". "most highly cited" (7.78%) and "top cited" (3.75%) papers are rarely mentioned. The Pearson $\chi^2$ test can be used to examine the extent to which the use of designations differs statistically significantly between the definitions formulated in the papers. When calculating the $\chi^2$ test for these data, it has to be taken into consideration that a paper can contain more than one designation. In this case Jann (2005) suggests calculating an $\chi^2$ test for each individual row in the table (here: for each designation). Repeating the test for each row makes it necessary to correct the level of significance; the (conservative) Bonferroni correction was used. For this correction, the significance level of $\alpha = .05$ is divided by the number of the repeated tests. As the results of the Pearson $\chi^2$ tests in Table 3 show, there is a statistically significant difference in the use of the designations "highly cited" and "most cited". The designation "highly cited" is used comparatively rarely (28.03%) in connection with a certain number of excellent papers (Top number). Instead, the designation "most cited" is used (50.96%).

There are different options (see above) where quantitative (statistical) procedures are used in a paper to identify excellent papers. The most frequent option, as shown in Table 1 and which Table 4 also makes clear, is to specify a certain number of top cited papers (57.55%). In around a quarter of the papers, the excellent papers are specified by percentile rank class (e.g. the class of the 10% most cited papers within a subject area and publication year). In around 17% of the papers a certain number of citations (such as 100 citations) is used to identify excellent papers. Finally in around 2% of the papers a certain distance from a mean (e.g. ten times the mean) in a reference set (e.g. the publications from a certain publication year and subject area) or, similarly, the method of characteristic score and scales (CSS) (Glänzel, 2007) is used to delimit excellent papers. As the breakdown of methods to identify excellent papers according to the definition of excellent papers in Table 4 shows, mainly percentile rank classes are used in (Thomson Reuters') literature databases (66.67%). Where a quantitative procedure is used ("Quantitative" column) a certain number of citations is used to determine excellent papers in 49% of the papers.

In Table 5 and in Table 6 characteristic values (mean values and most entries) are presented for the number of top papers, percentile rank class, and number of citations from the papers analysed in this study. For example, a mean citation rate is given in the tables for those studies which with the aid of a certain number of citations (e.g. papers with more than 100 citations) have identified excellent papers in a publication set. The objective of these analyses is to obtain information about the (average and most frequent) use of certain thresholds in all the papers researched, in all the papers in certain subject areas (see Table 5) and in various units under study (e.g. scientists, see Table 6). This information can be used as guide values for future bibliometric studies in order to identify excellent papers in various subject areas and in the units studied. As the figures for the number of top cited papers in Table 5 show, on average (arithmetic average) over 160 studies around 856 papers are used as the threshold for the selection of top cited papers. As this arithmetic average (and other arithmetic averages in the table) can be strongly affected by very high numbers in a few papers, the table also shows the median, which at 13.5 is much lower. On average, therefore, around 14 papers in a publication set are defined as top cited papers. This value varies between subject areas with 8 top cited papers for health sciences and 23 for multidisciplinary sciences (see Table 5). Across all the subject areas, the most frequently used value is 1 (the most highly cited paper) and second most frequently used is 10 (the ten most highly cited papers). As the results in Table 6 show, the information about top cited papers also fluctuates with the unit under study: from 10 top cited papers in an investigation of individual papers and 24 top cited papers in the investigation of scientists.

In addition to the characteristic values (mean values and most entries) for the top cited papers, Table 5 and Table 6 also show these values for percentile rank classes. As they show, a total of 63 papers gives an arithmetic average of 7.6 and a median of 3. This suggests that on average the authors of the papers use percentile rank classes which are (significantly) smaller than the top 10% most cited papers in their subject area and their publication (as they are used in the SCImago Institutions Ranking or in the Leiden Ranking). The most frequent percentile rank class which was selected in a total of 23 papers is the top 1% class; in 16 papers the top 10% class was chosen as the second most frequent. The differences in terms of the use of percentile rank classes between the individual subject areas (see Table 5) and units under study (see Table 6) are minor. These minor differences were expected as the percentile rank classes are a relative measure (and independent of the subject area and publication year), whereas the number of top cited papers is an absolute measure.

Finally, Table 5 and Table 6 show results on the number of citations which were used in the papers as thresholds for the identification of excellent papers. The figures in the table give the lower threshold; that means in the papers, the publications with a citation count higher than the threshold were designated for the identification of excellent papers. Over all the papers (n=47) the authors set an average threshold of 93 (arithmetic average) or 50 (median). As expected, this value varies considerably between the subject areas: It is highest in the life sciences at 189 citations (median) and lowest in the social sciences at 10 citations (median). For the units under study, the differences in the use of citation thresholds are much less pronounced (see the medians in Table 6). This result was also expected, as citations depend on the subject areas but hardly at all on the units (under study).

## 4 Discussion

As the subject of research excellence has received increasing attention (in science policy) over the last few decades, increasing numbers of bibliometric studies have been published dealing with excellent papers. However, many different methods have been used in these studies to identify excellent papers. The present quantitative analysis of the literature has been carried out in order to acquire an overview of these methods and an indication of an "average" or "most frequent" bibliometric practice. In around a third of the papers analysed in this study, a quantitative definition has been provided with which to identify excellent papers. With definitions which relate to an absolute number, either a certain number of top cited papers or papers with a minimum number of citations are selected. A definition based on an absolute number was used in most of the papers analysed in this study. Definitions based on a relative number should however be preferred to definitions based on an absolute number, as only they can be used for cross-field and cross-time-period comparisons. Vinkler (2012) says of one of these relative methods where a certain distance from the mean is used that it "may be preferably applied for obtaining the elite set. Accordingly, 3, 5 or 10 times the mean GF of the journals devoted to a field may serve as the lower limits for the highly cited papers" (p. 476, see also Kosmulski, 2013). However, only 2% of the papers included in this study and working with a quantitative method make references to the use of this method.

"One of the simplest way for determining the elite set of a set of papers assessed is to calculate the number (or share) of the papers in the top 0.01; 0.1; 1.0 or 10.0% papers within the total set" (Vinkler, 2012, p. 476). Among the papers which used a quantitative method to identify the excellent papers, around half worked with percentile rank classes. As it is also used in many cases by the database operators such as Thomson Reuters in the Web of Science, it appears to be <u>the</u> preferred method with which to identify an elite set (excellent papers) (see also Bornmann & Leydesdorff, 2013). The advantage of the method – compared to the mean-based method (see above) – is partly that it is possible to work with an expected value: For example, one can expect that with a random sample of percentiles in a database 1% of the publications will belong to the top 1%, 5% of the publications to the top 5% and so on. If these values deviate for the publications from an institution or a scientist (upwards or downwards) the performance will be better or worse than one can expect (Bornmann, 2013; Bornmann & Marx, 2013). On the other hand, no mean is calculated for citation counts with the use of percentiles (and percentile rank classes). The arithmetic mean should not be used with skewed count data.

However in the literature, very different percentile rank classes are used to identify excellent papers. Adams, Pendlebury, and Stembridge (2013), for example, work with the share of the papers in the world's top 1% most cited for that year. Levitt and Thelwall (2010) argue for the top 25%: "This study uses the $75^{th}$ percentile as its key indicator, because the mean citation can be highly skewed by a few very highly cited articles, the $90^{th}$ percentile sometimes covers too few articles, and the $50^{th}$ percentile is too crude an indicator because the numbers in some cases do not vary much" (p. 175). According to Albarran, Ortufno, and Ruiz-Castillo (2011) "the Leiden group has turned its attention to the upper tail of the distribution, and has introduced the percentage in the top 5% of the most highly cited papers as an indicator of scientific excellence" (p. 326). The results of this study agree strongly with the practice of the Leiden Group. Over a total of 63 papers, there is an arithmetic average of the top 7.6% (arithmetic average) or of the top 3% (median). The top 1% is used most frequently in the papers, followed by the top 10%.

With the thresholds presented in this study, in future, it will be possible to identify excellent papers based on an "average" or "most frequent" practice among bibliometricians. It is important to subject an "average" or "most frequent" practice in bibliometrics to an empirical examination, as there is no generally acceptable formula with which to identify scientific excellence or excellent papers. It is a normative definition (Leydesdorff, Bornmann, Mutz, & Opthof, 2011).

# Acknowledgements

The author is grateful to Loet Leydesdorff for exchanges of ideas.

Table 1. How were excellent papers defined in the study (in percent)?

| Definition | Subject area | | | | | Total (n=347) |
|---|---|---|---|---|---|---|
| | Life sciences (n=23) | Health sciences (n=30) | Physical sciences (n=72) | Social sciences (n=112) | Multidisciplinary sciences (n=110) | |
| Database | 17.39 | 0.00 | 5.56 | 0.89 | 10.91 | 6.05 |
| Qualitative | 21.74 | 23.33 | 11.11 | 20.54 | 23.64 | 19.88 |
| Quantitative | 8.70 | 16.67 | 36.11 | 25.00 | 35.45 | 28.82 |
| Top number | 52.17 | 60.00 | 47.22 | 53.57 | 30.00 | 45.24 |
| Total | 100.00 | 100.00 | 100.00 | 100.00 | 100.00 | 100.00 |

Notes. $\chi^2$ (df=12) = 37.22, p<.001; Cramér's V = .19 (small to medium effect size).
There was no information about the meaning of excellence in 18 papers.

Table 2. With which unit under study were the papers concerned when looking at excellent papers (in absolute numbers and in percent)?

| Unit under study | Absolute number | Percent |
| --- | --- | --- |
| Paper | 225 | 64.84 |
| Scientist | 54 | 15.56 |
| Journal | 50 | 14.41 |
| Other (e.g., patent) | 18 | 5.19 |
| Total | 347 | 100.00 |

Table 3. Which designations were used in the studies for excellence and what is the relationship of the designations to the definition of excellent papers (in percent)?

| Names | Definition of excellent papers | | | | |
|---|---|---|---|---|---|
| | Database | Qualitative | Quantitative | Top number | Total |
| Highly cited* | 85.71 | 69.57 | 69.00 | 28.03 | 51.59 |
| Most cited* | 19.05 | 20.29 | 15.00 | 50.96 | 32.56 |
| Most frequently cited | 4.76 | 7.25 | 4.00 | 16.56 | 10.37 |
| Most highly cited | 4.76 | 2.90 | 9.00 | 9.55 | 7.78 |
| Top cited | 4.76 | 0.00 | 7.00 | 3.18 | 3.75 |
| Much cited | 0.00 | 2.90 | 0.00 | 0.00 | 0.58 |
| Very highly cited | 0.00 | 0.00 | 1.00 | 0.00 | 0.29 |
| Total | 119.05 | 102.90 | 105.00 | 108.28 | 106.92 |
| Number of papers | 21 | 69 | 100 | 157 | 347 |
| Number of entries in the papers | 25 | 71 | 105 | 170 | 371 |

Note. * The difference in the designation of excellence between the various definitions is statistically significant (Pearson $\chi^2$ test, Bonferroni-adjusted $p < 0.05$).

Table 4. Which methods were used to identify excellent papers and how is the method related to the definition of excellent papers (in percent)?

| Method | Definition of excellent papers | | | |
| --- | --- | --- | --- | --- |
| | Database (n=21) | Quantitative (n=100) | Top number (n=157) | Total (n=278) |
| Number of top papers | 14.29 | 0.00 | 100.00 | 57.55 |
| Percentile rank class | 66.67 | 49.00 | 0.00 | 22.66 |
| Number of citations | 19.05 | 43.00 | 0.00 | 16.91 |
| Distance from mean | 0.00 | 6.00 | 0.00 | 2.16 |
| Number of co-citations | 0.00 | 2.00 | 0.00 | 0.72 |
| Total | 100.00 | 100.00 | 100.00 | 100.00 |

Table 5. Which number of top papers, percentile rank class, and numbers of citations do the authors give in the papers to identify excellent papers in various subject areas?

| Subject area | Number of papers | Arithmetic mean | Median | Most frequent entry | Second most frequent entry |
|---|---|---|---|---|---|
| **Number of top papers** | | | | | |
| Health sciences | 18 | 17.8 | 8.0 | | |
| Life sciences | 14 | 64.6 | 10.0 | | |
| Physical sciences | 34 | 329.1 | 20.0 | | |
| Social sciences | 61 | 1663.3 | 14.0 | | |
| Multidisciplinary sciences | 33 | 697.8 | 23.0 | | |
| Total | 160 | 855.7 | 13.5 | 1 (n=27) | 10 (n=21) |
| **Percentile rank class** | | | | | |
| Life sciences | 1 | 10.0 | 10.0 | 10 (n=1) | |
| Physical sciences | 15 | 8.0 | 1.0 | 1 (n=5) | 0.1 (n=3) |
| Social sciences | 6 | 7.7 | 5.5 | 10 (n=2) | 0.1 (n=2) |
| Multidisciplinary sciences | 41 | 7.4 | 3.0 | 1 (n=17) | 10 (n=11) |
| Total | 63 | 7.6 | 3.0 | 1 (n=23) | 10 (n=16) |
| **Number of citations** | | | | | |
| Health sciences | 5 | 32.6 | 24.0 | | |
| Life sciences | 3 | 229.7 | 189.0 | | |
| Physical sciences | 12 | 70.3 | 49.5 | | |
| Social sciences | 21 | 91.8 | 10.0 | | |
| Multidisciplinary sciences | 6 | 124.5 | 99.0 | | |
| Total | 47 | 93.0 | 50.0 | 99 (n=8) | 100 (n=5) |

Note. The most frequent and second most frequent entry is not given for the number of top papers and the number of citations for the individual subject areas. The variance of the values is too large and a clustering of single values is not given in order to extract individual values.

Table 6. Which numbers of top papers, percentile rank classes, and numbers of citations are given by the authors in the papers to identify excellent papers for various units under study?

| Unit under study | Number of papers | Arithmetic mean | Median | Most frequent | Second most frequent entry |
|---|---|---|---|---|---|
| **Number of top papers** | | | | | |
| Paper | 81 | 1514.1 | 10.0 | 1 (n=16) | 10 (n=10) |
| Scientist | 30 | 365.6 | 24.0 | 1 (n=4) | 10 (n=3) |
| Journal | 38 | 71.2 | 17.0 | 50 (n=6) | 1 (n=5) |
| Other | 11 | 54.2 | 10.0 | | |
| Total | 160 | 855.7 | 13.5 | 1 (n=27) | 10 (n=21) |
| **Percentile rank class** | | | | | |
| Paper | 44 | 9.1 | 5.0 | 1 (n=15) | 10 (n=12) |
| Scientist | 13 | 2.9 | 1.0 | 1 (n=7) | 0.1 (n=3) |
| Journal | 2 | 7.5 | 7.5 | | |
| Other | 4 | 6.5 | 7.5 | | |
| Total | 63 | 7.6 | 3.0 | 1 (n=23) | 10 (n=16) |
| **Number of citations** | | | | | |
| Paper | 39 | 101.9 | 50.0 | 99 (n=7) | 100 (n=5) |
| Scientist | 5 | 73.6 | 50.0 | | |
| Journal | 3 | 9.7 | 9.0 | | |
| Total | 47 | 93.0 | 50.0 | 99 (n=8) | 100 (n=5) |

Note. The most frequent and the second most frequent entry are not given if the case numbers are too low (n≤10) or the unit under study is in the category "Other".